\documentclass{article} 

\usepackage[utf8]{inputenc} 
\usepackage[english]{babel} 
\usepackage{amsmath}
\usepackage{amssymb}
\usepackage{txfonts}
\usepackage{mathdots}
\usepackage[classicReIm]{kpfonts}
\usepackage{graphicx}

\begin{document}


\noindent \textbf{Relationship between Cultural Values, Sense of Community and Trust and the Effect of Trust in Workplace }

\noindent Nazli Mohammad${}^{1}$, Yvonne Stedham${}^{1}$

\noindent ${}^{1}$College of Business, University of Nevada, Reno

\noindent \textbf{Abstract}

\noindent This paper provides a general overview of different perspectives and studies on trust, offers a definition of trust, and provides factors that play a substantial role in developing social trust, and shows from which perspectives it can be fostered. The results showed that trust is playing an important role in success for organizations involved in cross-national strategic partnerships. Trust can reduce transaction costs, promotes inter-organizational relationships, and improve subordinate relationships between managers.


\noindent 
\section{Introduction and Background}

\noindent The positive effects of social trust have been substantiated in several studies. Studies at the aggregate level---across nations or sub-national units---show that societies inhabited by more trustful citizens, experience higher levels of economic growth (Bj{\o}rnskov 2009; Knack and Keefer 1997), more effective democratic government (Knack 2002; Tavits 2006), and less dishonest behavior (Neville 2012). In short: social trust promotes desirable collective outcomes.

\noindent Increasingly, researchers from a variety of business disciplines are finding that trust can lower transaction costs, facilitate inter-organizational relationships, and enhance manager subordinate relationships. (Fukuyama, 1995). At the same time, we see a growing trend toward globalization in establishing alliances, managing, and hiring employees, and entering new markets. So, there is a need to view the concept of trust from the perspective of national culture.

\noindent Scholars in various disciplines have considered the causes, nature, and effects of trust. In this research, prior approaches to studying trust are considered, including the role of culture, the relation between belonging with a community and trust, and social characteristics. (Mayer and Davis, 1995) In this article, we address cultural factors that may have a direct effect on the level of trust in society. A definition of trust and key determinants of trust and outcomes are presented. The purpose of this article is to illuminate the key role of trust in critical social processes. Besides, given the importance of the behavioral consequences of trust, propositions for the relationship between trust and ethical judgment will be developed. Specific questions to be addressed are:

\noindent 1. What is trust? What role does culture play? How trust is established depends upon the societal norms and values that guide people's behavior and beliefs. Identifying factors influencing the development of trust.

\noindent  2. What are the importance and benefits of trust in the emerging global and multicultural workplace? How does trust develop and in what ways does national culture impact the trust-building process?

\noindent Exploring these questions is important for the following reasons:

\noindent Trust is critical to companies' and societies' success. Social relations are necessary to meet needs. When we learn more about trust across different cultures, more successful would be the interactions between people from different cultures. We know to which cultural norms and values facilitate or inhibit the formation of trust. Also, it helps in facilitating trust-building capacities between people. In short, knowing the differences in culture between the two societies can help avoid costly problems and failure. (Pitta, D. A., Fung, H. G., \& Isberg, S. 1999).

\noindent Trust enables people to take risks: where there is trust there is the feeling that others will not take advantage of me (Porter, Lawler, and Hackman, 1975). Trust is based on the expectation that one will find what is expected rather than what is feared (Deutsch, 1973). Thus, competence and responsibility are central to an understanding of trust (Barber, 1983; Cook \& Wall, 1980; Shapiro, 1990). Finally, trust encompasses not only people's beliefs about others but also their willingness to use that knowledge as the basis for action (Luhmann, 1979). Combining these ideas yields a definition of interpersonal trust as the extent to which a person is confident in and willing to act based on, the words, actions, and decisions of another. (McAllister, D.,1995).

\noindent 
\section{Determinants and Correlates of Trust}

\noindent The following section presents findings related to determinants and correlates of trust from the World Values Survey (\textit{WVS}). 

\noindent Society's level of economic development is closely linked with its level of interpersonal trust. Trust shapes economic development rates as well as the reverse. Studies show that rich societies have a much higher level of trust than poor ones.

\noindent In low-income countries, those with more education and those with postmaterialist values are only slightly higher on interpersonal trust than those with little or no education or those with materialist values. The 15 richest societies in the 1990 World Value Survey show that the more educated and the postmaterialists are much more likely to feel that most people can be trusted than are the less educated. So, within the low-income societies, the highly educated are only slightly higher on interpersonal trust, while among the 15 advanced industrial societies, the highly educated show a much higher level of interpersonal trust. Education does not seem to drive the process.

\noindent In addition, greater economic inequality means less trust. Less inequality and more racial/ethnic homogeneity have higher levels of trust. Then it is clear that economic inequalities weaken our civic life as a society. They are also less likely to have been materially successful.

\noindent Economic factors seem to play a major role in the emergence of interpersonal trust. But the result of this analysis suggests that a given society's religious heritage may be fully as important as its level of economic development in shaping interpersonal trust. Interpersonal trust shows remarkably strong linkage with the religious tradition of the given society and these religions were established long before the industrial development of these societies. The cross-national difference also reflects society's cultural heritage. Protestant and Confucian-influenced societies consistently show a higher level of Interpersonal trust than do historically Roman Catholic or Islamic societies.

\noindent Also, the World Value Survey data reveal a strong positive correlation between interpersonal trust and the functioning of democratic institutions. Interpersonal trust is strongly linked with democratic institutions, as well as economic development. One possible explanation of this configuration would be that economic development gives rise to democracy, which produces relatively high levels of Interpersonal trust. On the other hand, the results suggest that less hierarchical political institutions promote trust. (Martin Ljunge, 2014)

\noindent Politics, religion, family, education, and economics are functional for society.  Therefore, another important factor is considered social cohesion as levels of order and stability putting together by shared norms and values in society (Parsons 2013). These enable individuals to identify and contribute to common goals and share moral and behavioral norms that function as a base for interpersonal relationships. Larsen defines cohesion as the belief that citizens have in a given nation that shares a moral compass, which in turn provides a common ground for trust (Larsen 2013). It is then defined and measured by the number of individuals trusting each other to some degree (national identification and belief).

\noindent A cohesive society works towards the well-being of all its members, fights exclusion and marginalization, creates a sense of belonging, promotes trust. Characteristics of social cohesion are reciprocal loyalty and solidarity, strength of social relations and shared values, sense of belonging, trust among individuals of society and reduction of inequalities and exclusion.

\noindent There are three metrics to quantify social cohesion: individuals acting together towards a common goal, positive engagement around common goals, and a vulnerable and trusting attitude that fosters the sharing of private materials.

\noindent when it comes to distrust and diversity, most of the distrust is expressed by Whites who feel uncomfortable living amongst racial minorities. In other words, greater distrust may stem from prejudice rather than from diversity in itself (Grewal, 2016). Therefore, Putnam's conclusion that racial diversity leads to less altruism and cooperation amongst neighbors was incorrect. If there is a downside to diversity, it has less to do with the behavior of racial minorities and more to do with how Whites feel when living amongst non-Whites. Therefore, it may be the instability of diverse communities, rather than diversity itself, that erodes trust.

\noindent  On the other hand, common race and sex had little effect on sharing or inferred trust (Matrinovich, 2017). This finding suggests that emotional expression may play a more powerful role in resource sharing than even race or sex, said Tsai, director of Stanford's Culture and Emotion Lab. 

\noindent Much of the research on the explicit relationship between culture and trust, employs the Hofstede (1980) cultural Framework. The concept of cultural dimension is a specific aspect of a culture that can be measured relative to other cultures (Hofstede). Six such cultural dimensions (CDs) have been identified: power distance, collectivism versus individualism, femininity versus masculinity, uncertainty avoidance, long-term versus short-term orientation, and indulgence versus restraint. These dimensions operate together, and their operation is influenced by political and economic circumstances and by individual personality factors. The conceptual domain Clark labels relation to self encompasses issues of personality and self-concept (1990). 

\noindent Two of Hofstede's (1984) dimensions deal with relation to self. First, individualism/collectivism refers to the relationship between the individual and the collectivity. Norms and values associated with individualism/collectivism reflect the way people interact, such as the importance of unilateral versus group goals, the strength of interpersonal ties and trust, respect for individual accomplishment, and tolerance of individual opinion. Second, masculinity/femininity concerns the dominant values in society (Singh, 1990). The masculinity/ femininity dimension assesses the degree to which tough values, such as assertiveness, success, and competition, prevail over tender values, such as nurturance, service, and solidarity. 

\noindent In relating the cultural dimensions to trust, it has been helpful to consider the scientific literature that describes five distinct cognitive patterns that can be employed by the trustor when forming such expectations: calculative, capability, prediction, intentionality, and transference. These cognitive patterns are called cognitive trust-building processes (CTBPs).

\noindent The best device for creating trust is to establish and support trustworthiness. Factors of perceived trustworthiness are Ability, Benevolence, and Integrity. Both ability and integrity belong to the cognitive dimension of trust. Benevolence, on the other hand, belongs to the affective dimension. Benevolence is the extent to which a person believes that other parties will do well to him or her, aside from egocentric profit motive (Mayer et al., 1995).

\noindent 
\section{Community }

\noindent Based on existing research, cohesion in a community positively affects trust. Social cohesion refers to the extent of connectedness and solidarity among groups in society that enhances trust. In contrast, therefore, social isolation would be associated with less trust. Social isolation is a complex issue linked to physical, emotional, and psychological well-being, and influenced by personal, community, societal factors. Social isolation impacts the overall quality of life and has also been linked to depression, anxiety, social stigma, dementia, and increased risk of cognitive decline Sense of community.

\noindent Low-income people and seniors are among the most vulnerable to social isolation. Poverty and low income have both been found to increase the risk of loneliness and social isolation. Certain characteristics place people, particularly older adults, at greater risk of becoming socially isolated including having a low income; living alone; poor health; not having children or contact with family; and lack of access to transportation. 

\noindent Seniors are at greater risk of becoming lonely and socially isolated, largely due to factors that compound to limit social contacts, such as declining income, mobility issues, and the death of friends and family.

\noindent Not only senior people are at risk of social isolation, but also young people are prone to loneliness. The uncertain political situation, the lack of leadership in the country, a high incidence of child abuse, rape, sexual abuse, incest, and violence are the factors that increase loneliness among young people (Lch doman; A le Roux, 2010). 

\noindent Other factors that cause and contribute to loneliness are mentioned in the following:

\noindent \underbar{Personality:} anxiety, an inability to assert oneself and hypersensitivity, poor self-concept, shyness are other aspects of the personality that contribute to loneliness. Isolated people are also more emotionally mature, inclined to feel guilty, and less sociable than others.

\noindent \underbar{Depression:} people with poor social skills are more inclined to become depressed and yet depression is inclined to loneliness. A lack of feelings of belonging, support, and intimacy caused by poor communication promotes loneliness. Social estrangement and separation and rejection also contribute to loneliness (Lch doman; A le Roux, 2010).

\noindent \underbar{Marital status:} Married people usually feel less lonely.  Life changes include the abandonment of the parental home, a new occupation, the separation of loved ones, the termination of an important relationship, and relocation increase loneliness (Lch doman; A le Roux, 2010). 

\noindent \underbar{Problems in relationships:} A lack of feelings of belonging, support, and intimacy caused by poor communication promote loneliness. Social estrangement and separation and rejection also contribute to loneliness.

\noindent  \underbar{Illness or physical disability:} this may lead to limited contact with other people. Research indicates that HIV/AIDS and eating disorders isolate sufferers from social contact (Lch doman; A le Roux, 2010).

\noindent \underbar{Religious faith:} it correlates negatively with loneliness. However, religions with strict behavioral prescriptions are likely to isolate individuals from free social interaction. This could result in loneliness.

\noindent \underbar{Academic achievement:} Research indicates that students whose academic performance is poor are more likely to be lonely.

\noindent Living alone, advancing of age, gender, level of education, place of residence, visual and hearing impairments, widowhood, co-morbid illness, and depression have been found highly significantly associated with loneliness in older adults in previous studies.

\noindent 
\section{Build a Culture of Trust in the workplace}

\noindent Trust allows humans to form communities, cooperate and even, at times, find solutions beyond plain self-interest. Trust affects how we form relationships with our family and friends, and why and how we develop business relationships or decide to buy products in the marketplace (Cook et al., 2009; Sztompka, 2000; Zucker, 1986). 

\noindent The importance of trust has been cited in organizational studies areas as communication, leadership, management by objectives, negotiation, game theory, performance appraisal, labor-management relations, and implementation of self-managed work teams. Working together often involves interdependence, and people must therefore depend on others in various ways to accomplish their personal and organizational goals.

\noindent PWC's 2016 Global CEO Survey revealed that 50\% of CEOs worldwide consider a lack of trust to be a major threat to their organizational growth. Paul J Zak has invested decades researching the neurological connection between trust, leadership, and organizational performance. 

\noindent Zak discovered that compared with people at low-trust companies, people at high-trust companies report 74\% less stress, 106\% more energy at work, 50\% higher productivity, 13\% fewer sick days, 76\% more engagement, 29\% more satisfaction with their lives, and 40\% less burnout. He also found a direct link between oxytocin levels and empathy which is essential for creating trust-based relationships and trust-based organizations. The higher the oxytocin, the higher the empathy. The higher the empathy, the deeper the connection. Zak's research found that this level of vulnerability increases trust and cooperation in others because it stimulates oxytocin production.

\noindent The importance and benefits of trust, and the emerging global and multicultural workplace, highlight the need for us to understand how trust develops and the ways national culture impacts the trust-building process. Although trust may form in a variety of ways, whether and how trust is established depending upon the societal norms and values that guide people's behavior and beliefs (e.g., Hofstede, 1980).

\noindent Since each culture's collective programming results in different norms and values, the processes trustors use to decide whether and whom to trust may be heavily dependent upon a society's culture. Indeed, one of the greatest impacts of culture is on how information is used to make decisions (Triandis, 1972).

\noindent 
\section{Conclusion}

\noindent The concept of trust is playing an increasing role in success for the organizations that become involved in cross-national strategic partnerships. So, understanding the relationship between culture and trust is necessary. Trust can reduce transaction costs, promote inter-organizational relationships, and improve subordinate relationships between managers. Series of research propositions demonstrate how societal norms and values influence the application of the trust-building processes. Finally, individual personality and organizational culture surely affect trust forming.

\noindent 
\section{}

\noindent 
\section{References: }

\noindent Cook, K. S., Levi, M., \& Hardin, R. (Eds.). (2009). Whom can we trust? How groups, networks, and institutions make trust possible. Russell Sage Foundation.

\noindent Ethnic Diversity in the Neighborhood and Social Trust of Immigrants and Natives. A Replication of the Putnam (2007) Study in a West-European Country.

\noindent Mayer, R. C., Davis, J. H., \& Schoorman, F. D. (2006). An integrative model of organizational trust. Organizational trust: A reader, 82-108.

\noindent Mayer, Roger C., et al. ``An Integrative Model of Organizational Trust.'' The Academy of Management Review, vol. 20, no. 3, 1995, pp. 709--734. JSTOR.

\noindent McAllister, Daniel J. ``Affect- and Cognition-Based Trust as Foundations for Interpersonal Cooperation in Organizations.'' The Academy of Management Journal, vol. 38, no. 1, 1995, pp. 24--59. JSTOR

\noindent McCauley, D. P., \& Kuhnert, K. W. (1992). A theoretical review and empirical investigation of employee trust in management. Public Administration Quarterly, 265-284.

\noindent McMahon, J.M., Harvey, R.J. The Effect of Moral Intensity on Ethical Judgment. J Bus Ethics 72, 335--357 (2007).

\noindent Mohammad, N.; Stedham, Y.; (2020). Measuring Trust and its Role in Workplace, Proceedings of GSA Fall 2020 Symposium.

\noindent Mohammad, N.; Stedham, Y.; (2021). Cultural Values and Sense of Community Influences on Trust, Proceedings of Academy of Business Research Summer 2021 Conference.

\noindent Mohammad, N.; Stedham, Y.; (2021). The power of Trust. Proceedings of Three Minute Thesis (3MT) 2021 Competition.

\noindent Nazli Mohammad, Yvonne Stedham. Relationship between Cultural Values, Sense of Community and Trust and the Effect of Trust in Workplace. 2021. $\mathrm{\langle }$hal-03257439$\mathrm{\rangle }$.

\noindent Porter, L. W., Lawler, E. E., and Hackman, R., (1975), "Behaviour in Organizations", New York: McGraw-Hill.

\noindent Reidenbach, R.E., Robin, D.P. Toward the development of a multidimensional scale for improving evaluations of Business Ethics. J Bus Ethics 9, 639--653 (1990).

\noindent Rokach, A., Orzeck, T., Moya, M.C. \& Exposito, F. 2002. Causes of loneliness in North America and Spain. European Psychologist, 7:70-79.

\noindent Rothstein, Bo (2005), Social Traps and the Problem of Trust, Cambridge, UK: Cambridge University Press.

\noindent Schoorman, F. D., Mayer, R. C., \& Davis, J. H. 1996b. Organizational trust: Philosophical perspectives and conceptual definitions. Academy of Management Review, 21: 337--340.

\noindent Zak P. J, The Neuroscience of Trust: Management behaviors that foster employee engagement, issue (pp.84--90) of Harvard Business Review, 2017.

\noindent Zucker, L. G. (1986). Production of trust: Institutional sources of economic structure, 1840--1920. Research in organizational behavior.

\noindent 

\begin{enumerate}
\item  Top of Form
\end{enumerate}

\noindent 

\noindent 

\noindent 

\noindent 

\noindent 

\noindent 

\noindent

\end{document}